\documentclass[nobibnotes,10pt,twocolumn,prl,aps,floatfix,superscriptaddress,longbibliography]{revtex4-1}

\usepackage{graphicx}
\usepackage{xcolor}
\usepackage{amsthm,amsmath,amssymb}
\usepackage[english]{babel}
\usepackage{siunitx}
\usepackage[sort&compress]{natbib}

\renewcommand{\vec}[1]{\boldsymbol{#1}}

\begin{document}

\title{Dissipation in mesoscale superfluids}

\author{Adrian Del Maestro}
\email{Adrian.DelMaestro@uvm.edu}
\affiliation{Department of Physics, University of Vermont, Burlington, VT 05405, USA}

\author{Bernd Rosenow}
\affiliation{Institut f\"ur Theoretische Physik, Universit\"at Leipzig, 
D-04103, Leipzig, Germany}

\begin{abstract}
We investigate the maximum speed at which a driven superfluid can flow through
a narrow constriction with a size on the order of the healing length.
Considering dissipation via the thermal nucleation of quantized vortices, we
calculate the critical velocity for superfluid $^4$He and ultracold atomtronic
circuits, identify fundamental length and velocity scales, and are thus able to
present results obtained in widely different temperature and density ranges in
a universal framework. For ultra-narrow channels we predict a drastic reduction
in the critical velocity as the energy barrier for flow reducing thermally
activated phase slip fluctuations is suppressed. 
\end{abstract} 

\maketitle

% =============================================================================== 
% Introduction
% =============================================================================== 

The flow of dissipationless atomic supercurrents in neutral superfluids is one
of the most dramatic manifestations of macroscopic quantum coherence
\cite{anderson,varoquauxrmp,chien}, with applications to matter wave 
interferometry \cite{sato,Schumm+05,Chiow+11}.  Recently, there has been
increased interest in dimensionally confined superfluids, due to progress
in manufacturing nanoscale channels and fountain effect devices 
for studying the flow of superfluid helium
\cite{Hoskinson:2005km,hoskinson,Cheng:2015yh,beamish:2016ay,Vekhov:2014ym,
Vekhov:2012mf,Ray:2010mf,Ray:2008ou,Boninsegni:2007ll,Savard:2011hs,Savard:2009fc,duc, 
Velasco:2012de,velasco,Botimer:2016pd,Fil:2009bpa,DelMaestro:2011ll,Pollet:2014tq}
and the availability of trapped non-equilibrium atomic Bose-Einstein condensates
\cite{raman,watanabe,neely,ramanathan,lee,wright,Murray:2013br,
Wright:2013bg,eckel,Jendrzejewski:2014dy,mathey,weimer,husmann,singh,Li:2016ux}.
Common to these experiments in vastly different density and interaction regimes
is an observed increase in dissipation for highly confined systems.

In general, superflow is possible at speeds less than a superfluid critical
velocity set by the Landau criterion  $v_c \le \min \varepsilon(p)/p$ below
which there are no accessible excitations $\varepsilon(p)$ with momentum $p$
\cite{landau}. Among the different types of excitations in superfluids,
quantized vortices \cite{anderson,iordanskii, feynman,langer,volovik} give rise
to the smallest $v_c$.  For flow through a cylindrical channel, if the total
kinetic energy is converted into vortex rings with the size $a$ of the
constriction, the Landau criterion predicts a critical velocity $v_{c,F} \sim
(\kappa/a)\ln(a/\xi_0)$ \cite{feynman} where $\kappa = h/m$ is the quantum of
circulation for condensed bosons of mass $m$ and $\xi_0$ is a characteristic
length scale of the superfluid. This prediction (due to Feynman) has been born
out by nearly a half-century worth of superfluid massflow observations with
\emph{temperature independent} critical velocities \cite{varoquauxrmp,
harrison}.  However,  it must ultimately break down as the constriction radius
approaches $\xi_0$. Moreover, any observed temperature dependence of $v_c$ can
only be described by the existence of an energy activation barrier for the
creation of vortices.

In this letter, we consider confined mesoscale superflow  through
quasi-one-dimensional (1d) constrictions with a characteristic size $a$
approaching the temperature $(T)$ dependent correlation (healing) length
$\xi(T)$, and find a strong increase in dissipation when $a/\xi(T)$ approaches
one.  Going beyond previous studies \cite{varoquauxrmp,weimer}, we (i)
quantitatively predict  the temperature, size, and drive dependence of the
critical velocity without adjustable parameters, (ii) use a paradigmatic
orifice geometry to model the enhancement of vortex creation in spatially
inhomogeneous flow near a sharp boundary, which significantly  lowers critical
velocities, (iii) point out the universality between high density $^4$He
\cite{hoskinson,duc} and low density atomic condensates
\cite{neely,ramanathan,raman,weimer}, by characterizing constrictions via the
dimensionless length $a/\xi$ and measuring velocities in units of $ v_0 =
{\kappa}/ \left({4 \pi \xi_0}\right)$, and (iv) describe the crossover to the
purely 1d limit, a Luttinger liquid in the thermal regime.
Predictions are expected to be logarithmically accurate in the
critical regime while corrections of order unity may arise when extrapolating
to lower $T$.

%
% ------------------------------------------------------------------------------- 
\begin{figure}[t]
\begin{center}
\includegraphics[width=0.8\columnwidth]{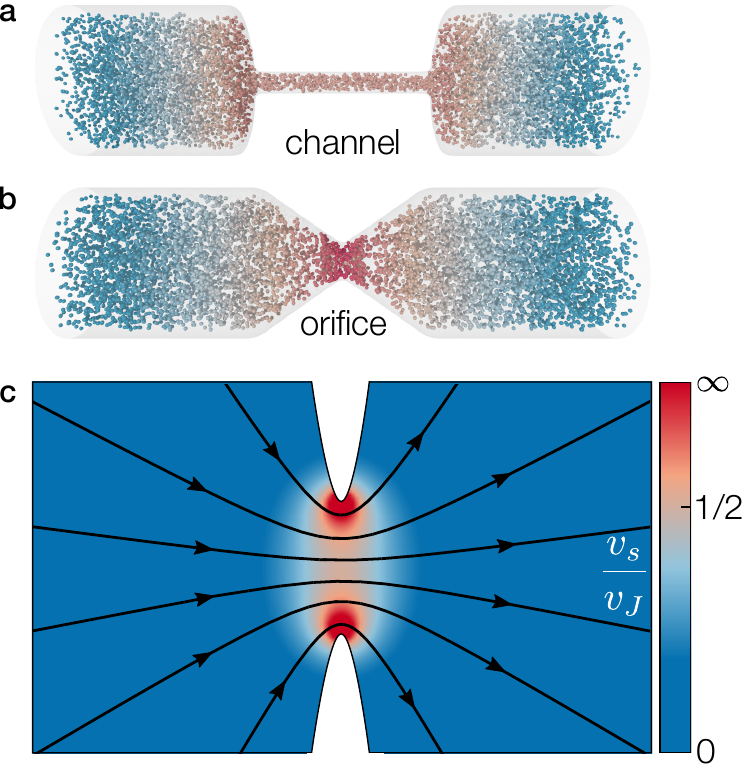} 
\end{center}
\caption{\textbf{a} A current of superfluid atoms driven
    through a long channel with a homogeneous (radius independent) flow
    profile. \textbf{b} Flow through a narrow orifice formed from the surface
    of revolution of a hyperbola around the flow axis.  \textbf{c} Velocity
    field $\vec{v}_s$ for the potential flow though an orifice in units of the
    average flow speed $v_J$.  The flow
direction is indicated by black lines, with the magnitude 
diverging as a power law at the orifice boundary  \cite{supplemental}.
\label{flowprofiles.fig}
}
\end{figure}
% ------------------------------------------------------------------------------- 
%

% =============================================================================== 
% Methods
% =============================================================================== 

\begin{figure}[t] \begin{center}
\includegraphics[width=0.75\columnwidth]{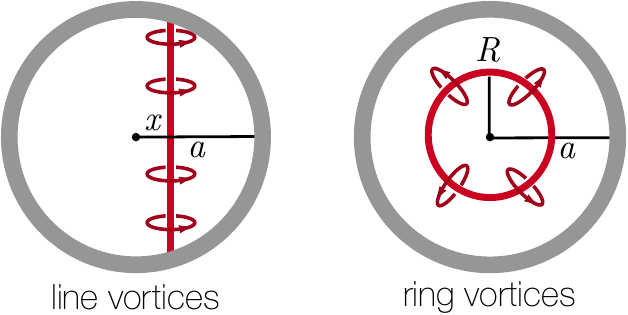} \end{center}
\caption{Two vortex types that can be nucleated in a confined geometry. Left: A
    line vortex located a distance $x$ from the center of the orifice or
    channel.  The vortex line begins and ends on the boundary, with the flow
    circulating around it.  Right: A vortex line can detach from the boundary
    and close on itself forming a vortex ring of radius $R$.  Arrows indicate
    the circulation of quantized flow around the vortex core.}
\label{vortextypes.fig} 
\end{figure}

We begin by considering superflow between reservoirs with a
chemical potential difference $\Delta \mu$  (pressure difference $\Delta P$)
connected by prototypical geometric constrictions with  
either a ``channel'' or ``orifice'' shape as seen in Fig.~\ref{flowprofiles.fig}.
Channel flow is spatially homogeneous with a constant superfluid velocity $v_s$
that is representative of flow through long narrow cylindrical pores.  Flow
through an orifice can be studied by considering a hyperbolic surface of
revolution connecting two bulk reservoirs where the potential flow problem can
be solved analytically \cite{supplemental}.  The solution 
is characterized by divergent flow near the boundaries
as seen in Fig.\ref{flowprofiles.fig}c, where the creation of line vortices
(Fig.~\ref{vortextypes.fig} left) is facilitated.  Ring vortices in the
center of the orifice (Fig.~\ref{vortextypes.fig} right) are not strongly
affected by the spatial dependence of flow near the boundaries.
%
% ------------------------------------------------------------------------------- 
% ------------------------------------------------------------------------------- 
%

The energy cost for creating a quantized vortex is due to (i) the kinetic
energy of circular superflow around the core region, and (ii) the loss of
condensation energy within the core \cite{iordanskii}. For a vortex ring of
radius $R$ (length $\mathcal{L} = 2\pi R$) or a line vortex (length
$\mathcal{L}$) in a constriction of radius $a$,  the combination of these
effects yields 
%
%*********************************  vortex ring energy  ********************
\begin{equation}
E_{\rm tension} =  \frac{\mathcal{L} }{4\pi} \rho_s \kappa^2 
\left[ \ln \frac{\ell}{\xi}  +  \alpha \right] 
\label{eq:Etension}
\end{equation}
%*************************************************************************
%
where $\ell = a$ for line vortices, $\ell = R$ for ring vortices, and the
constant $\alpha$ depends of the vortex type and model of the core.  Solving
the Gross-Pitaevski (GP) equation, one finds that $\alpha = 0.385$ for line vortices
\cite{Pi61} and $\alpha = \ln 8 - 2 + 0.385 \simeq 0.464$ for ring vortices
\cite{RoGr71}.  Obtaining a more accurate model of the vortex
core is possible via numerical simulations in low \cite{Wi99} and high
density superfluids \cite{OrCe95,GaReRo14}.   The results are consistent with
the GP value of $\alpha$ and show only only weak density dependence.  The
reduction of the kinetic energy of superflow due to interaction with the vortex
is 
%
%***********************   flow energy ******************
\begin{equation}
E_{\rm flow} =  \kappa \rho_s \iint \vec{v}_s \cdot d\vec{S}  \ \ , 
\label{eq:Eflow}
\end{equation}
%************************************************
%
where the integral is over the area bounded by the vortex ring, or between the
vortex line and boundary.  The total energy is $E = E_{\rm tension} + E_{\rm
flow}$.  To unify the description of driven quantum fluids, we employ the Josephson
relation in 3d \cite{josephson} 
%
%************************ Josephson scaling relation **************
$\kappa^2 \rho_s(T) \xi(T) = 4 \pi^2 k_{\rm B} T_c$
%*****************************************************
%
where $\rho_s(T)$ is the superfluid mass density for a transition temperature
$T_c$ and $\xi(T) = \xi_0(1-T/T_c)^{-\nu}$ and $\nu$ is the correlation length
critical exponent.
We numerically checked that the Josephson relation is 
valid to within $20\%$ down to $T/T_c \approx 0.7$, for details see
\cite{supplemental}. For flow through an orifice with speed $v_J$,  we obtain the energy
barrier for a line vortex located a distance $x$ from its center:
\begin{align}
    \beta_c E_{\rm line}(x)  &=  2  \pi  \frac{a}{\xi} \sqrt{1 - \left(\frac{x}{a}\right)^2}\left[\ln\left(1 - \frac{|x|}{a}\right) 
    + \ln \frac{a}{\xi} + \alpha \right] \nonumber \\
    & \qquad
    - \frac{v_J}{v_0} \left( \frac{a}{\xi} \right)^2 \frac{\pi^2 \xi}{2 \xi_0}
    \left(1-\frac{x}{a}\right)  .
\label{eq:betaELineOrifice}
\end{align}
and that for a centered ring vortex with radius $R$: 
\begin{align}
    \beta_c E_{\rm ring}(R) &= 2 \pi^2 \frac{R}{\xi} \left( \ln \frac{R}{\xi} +
\alpha\right) \nonumber \\
& \quad    - \frac{v_J}{v_0} \left(\frac{a}{\xi}\right)^2 \frac{\pi^2 \xi}{\xi_0} 
    \left[1-\sqrt{1-\left(\frac{R}{a}\right)^2}\, \right] 
\label{eq:betaERingOrifice}
\end{align}
where $\beta_c = 1/(k_{\rm B} T_c)$.  From these expressions (and those for 
channel flow derived in the supplementary material \cite{supplemental})
we observe the emergence of natural length ($\xi_0$) and velocity $(v_0 =
\kappa/(4\pi\xi_0))$ scales that are essential for constructing a universal
theory of dissipative superfluids.  

The velocity of superflow at finite $T$ is limited by the thermal nucleation of
quantized vortices which traverse the flow lines leading to a change of phase
of the superfluid order parameter by $\pm 2\pi$, (see e.g.
Ref.~[\onlinecite{langernuc}]). The decay of a persistent current is then
governed by vortex energetics via: 
\begin{equation}
\Gamma = \Gamma_0 \left[ \mathrm{e}^{-\beta E_{\rm max}(+v_J)} - \mathrm{e}^{-\beta E_{\rm max}(-v_J)} \right] \  ,
\label{eq:decayRate}
\end{equation}
where $h\Gamma = \Delta \mu = m \Delta P/\rho_s$ drives total mass flow
$J = \rho_s \iint \vec{v}_s \cdot d \vec{S} \equiv \pi a^2 \rho_s v_J$ and 
$
E_{\rm max} \equiv \max_{\mathcal{L}} E 
$
is the saddle point of the vortex excitation energy over the domain
of the channel or orifice.  The difference of rates in Eq.~\eqref{eq:decayRate}
corresponds to the contributions from vortices which reduce and increase the
superflow, respectively.  The attempt rate $\Gamma_0$ is related to the phase
space available for vortex excitations and is geometry dependent:
\begin{equation}
    \Gamma_0 = \frac{1}{\tau_{GL}} \frac{L a}{\xi^2} \left\{ 
\begin{array}{cl}
    \frac{\pi a}{\xi} & {\rm vortex\ ring} \\
2 \pi & {\rm vortex \ line}
\end{array}
\right. ,  \label{attemptrate.eq}
\end{equation}
with $\tau_{\rm GL}^{-1} = 16 k_{\rm B}(T_c -T)/h$ the Ginzburg-Landau
relaxation rate.  The decay rate in Eq.~(\ref{eq:decayRate}) contains the main
contribution from integration over zero modes,  corresponding to a translation
of the vortex with action $S_{v}$ in both time and space. In
addition, there are Jacobians due to the change of coordinates from the
superfluid phase $\Phi$ to the radius $R$ of the vortex ring or the location
$x$ of the vortex line, and a contribution from integration over the negative
eigenvalue mode.  As previously discussed \cite{coleman1, *coleman2,schulman},
the Jacobian is $\sqrt{S_{v}/2 \pi}$  for the zero modes, and the
negative eigenvalue mode contributes a factor of similar magnitude.  We have
verified that the combination of all these factors is of order unity and
thus neglect them. Other modifications to the
pre-factor in Eq.~\eqref{eq:decayRate} could result when considering the
microscopic details of dynamics and vortex evolution inside the constriction
\cite{Bulgac:2011ee, Yefsah:2013bl, Ku:2014ip, Donadello:2014cz,
Serafini:2015fk, Ku:2016fk} and would introduce quantitative logarithmic
corrections to the nucleation theory.

Evaluation of the critical velocity from Eq.~\eqref{eq:decayRate} proceeds as
follows.  For a given flow profile and vortex type, we maximize the 
vortex energy as a function of $\pm v_J$ over the spatial domain of possible
configurations.  This leads to critical vortex positions $x^\ast(\pm v_J) \in
(-a,a)$ for line vortices and critical radii $R^\ast(\pm v_J) \in (0,a)$ for
ring vortices.  Vortices with a length smaller than the critical one will tend
to collapse, or anihilate at boundaries, while those larger can proliferate,
leading to dissipation.  For a fixed constriction
radius $a/\xi_0$, temperature $T/T_c$ and external drive potential
$\Gamma_0/\Gamma$, Eq.~\eqref{eq:decayRate} can be numerically solved
self-consistently giving the critical velocity when $v_J = v_c$.  

When $a/\xi_0 \gg 1$, the boundaries of the constriction no longer play an
important role and only uniform channel flow is relevant.  Due to
the resulting large critical velocities, energy increasing fluctuations
can be neglected and the saddle point energies can be
found analytically.  The resulting pair of transcendental equations
\begin{align}
\label{eq:RcritEqn}
\frac{R^\ast}{\xi} &= \frac{1}{\pi^2}\frac{T}{T_c} \frac{\ln \frac{\Gamma_0}{\Gamma}}
    {\ln \frac{R^\ast}{\xi} + \alpha - 1 } \\
\label{eq:vcEqn}
\frac{v_c}{v_0} &= \frac{\xi_0}{R^\ast} \left(\ln \frac{R^\ast}{\xi} +
    \alpha + 1\right) 
\end{align}
can be solved numerically for $R^\ast$ and $v_c$ \cite{supplemental}.  

\begin{figure}[t]
\begin{center}
\includegraphics[width=\columnwidth]{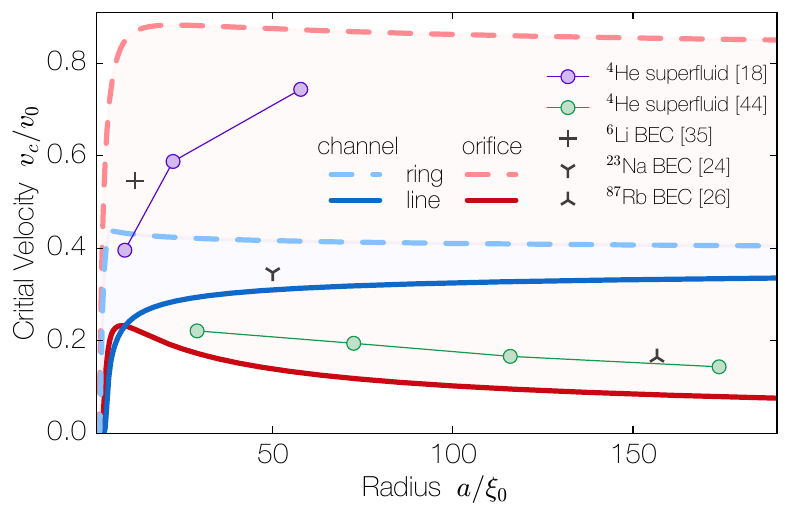}
\end{center}
\caption{ Upper and lower bounds on the critical velocity are indicated by
    lines from the channel ($L=10^3\xi_0$, blue) and orifice ($L=10\xi_0$, red) flow
profiles for both ring (dashed) and line vortices (solid) at 
$T=0.7T_c$, $\hbar \Gamma = 0.1 k_B T_c$ and $\nu \simeq 0.6717$.  Symbols show
experimental massflow results for superfluid helium \cite{duc,harrison} and
Bose-Einstein condensates \cite{neely,raman,weimer}. Details are provided in
the supplemental information \cite{supplemental} where we discuss the
relationship between the zero temperature healing length and $\xi_0$ in the weakly interacting
Bose gas \cite{Prokofev:2004js}. }
\label{fig:exp_comparison_bec}
\end{figure}

Our main results for the critical velocity in both the channel and orifice flow
profiles are shown as lines in Fig.~\ref{fig:exp_comparison_bec}.  When
employing the scales $v_0$ and $\xi_0$, mass flow measurements in drastically
different density, interaction, and temperature regimes are well bounded by the
vortex nucleation theory, and experiments on confined superfluid $^4$He and
low-dimensional Bose-Einstein condensates can be directly compared.
%
% ------------------------------------------------------------------------------- 
% ------------------------------------------------------------------------------- 
%
For both flow profiles, line vortices have lower activation energies than ring
vortices, giving smaller velocities and a lower bound on $v_c$.  An absolute upper
bound is provided by ring vortices in the orifice flow profile.  In the limit
$a \gg \xi_0$, where the geometry approaches that of bulk flow, we recover the
intrinsic superfluid velocity due to the nucleation of vortex rings analyzed by
Langer and Fisher \cite{langer}. For $^4$He, we have used a vortex core size
determined from specific heat measurements \cite{SiAh84,supplemental}, and have
thus been able to correct a long-standing inconsistency of 27 orders of
magnitude in the applied pressure difference employed in
Ref.~[\onlinecite{langer}] to obtain agreement with experiments.  For tight
constrictions, both the lower and upper bound turn towards smaller velocities
indicative of enhanced dissipation. 

While details of additional experiments are discussed in the supplemental
material \cite{supplemental}, Fig.~\ref{fig:exp_comparison_bec} includes data
from two $^4$He studies whose critical velocities display a clear temperature
dependence -- a signature of activated behavior.  Harrison \emph{et al.}
\cite{harrison} measured flow in networks of imperfect pores of varying radii
which should provide a lower bound on flow speeds, behavior consistent with our
results.  Recent measurements on single nanopores by Duc \emph{et al.}
\cite{duc} exhibit drastically different behavior: large critical velocities
that \emph{decrease} as the radius approaches the correlation length.  While
the microscopic details of the flow profiles are not known, $v_c$ for the
narrowest pore is bounded by the channel prediction, consistent with the
reported aspect ratio of 10:1 and suggesting a crossover to strongly
dissipative quasi-1d flow.

Fig.~\ref{fig:exp_comparison_bec} also includes results from analogous
neutral atomtronic circuits using utracold bosonic and fermionic condensates
\cite{chien}.   Here, the ``orifice'' can be replaced with a quantum point
contact between two resonantly coupled Fermi gases \cite{husmann} or
channel-like flow can be driven by the discharge of an bosonic atom capacitor
\cite{lee} or by dragging an optical potential through a simply
\cite{neely,weimer} or multiply connected \cite{ramanathan,wright,eckel}
Bose-Einstein condensate.  For the latter, our nucleation theory yields
$v_c \approx 100~{\rm \mu m/s}$ for the drag and $v_c \approx 1~\text{mm/s}$
for the toroidal flow in remarkable agreement with measurements and more
microscopic theoretical analysis \cite{crescimanno,eckel,mathey}.

Intuition for the dissipation in narrow pores with radius $a \approx \xi$ can
be gained by considering the unoptimized energy of line vortices with position
$x=0$ at the center of the channel.  This approximation is qualitatively
correct since narrow pores can be expected to be in the channel flow regime
with line vortex activation energies comparable to temperature. The vortex
energy is found from the sum of Eqs.~(\ref{eq:Etension}) and (\ref{eq:Eflow})
with $x=0$:
\begin{equation}
    \beta_c E_{\rm line}(x=0) = 2\pi \frac{a}{\xi} \left(\ln \frac{a}{\xi} +
    \alpha\right) - \frac{\pi^2 \xi}{2 \xi_0} \frac{v_s}{v_0}
    \left(\frac{a}{\xi}\right)^2  \  , 
\label{eq:Ex0Line}
\end{equation}
with resulting critical velocity
\begin{align}
    \frac{v_c}{v_0} &= \frac{2}{\pi^2} \frac{\xi \xi_0}{a^2} 
    \frac{T}{T_c}   \sinh^{-1} \left\{ \frac{1}{64\pi} \frac{\xi^2}{a L}
        \left(1-\frac{T}{T_c}\right)^{-1}
         \right. \notag \\
&\qquad \times \left. 
\exp\left[ \frac{2 \pi a}{\xi} \frac{T_c}{T} 
\left(\ln \frac{a}{\xi} + \alpha  \right)\right]\right\} \ \ .
\label{eq:vcNoopt}
\end{align}
As $a/\xi \to 1^{+}$, the activation energy in the exponent of
Eq.~\eqref{eq:vcNoopt} decreases rapidly, and the small multiplier of the
exponential is no longer compensated. As a consequence, the critical velocity
drops by several orders of magnitude. 

\begin{figure}[t]
\begin{center}
\includegraphics[width=\columnwidth]{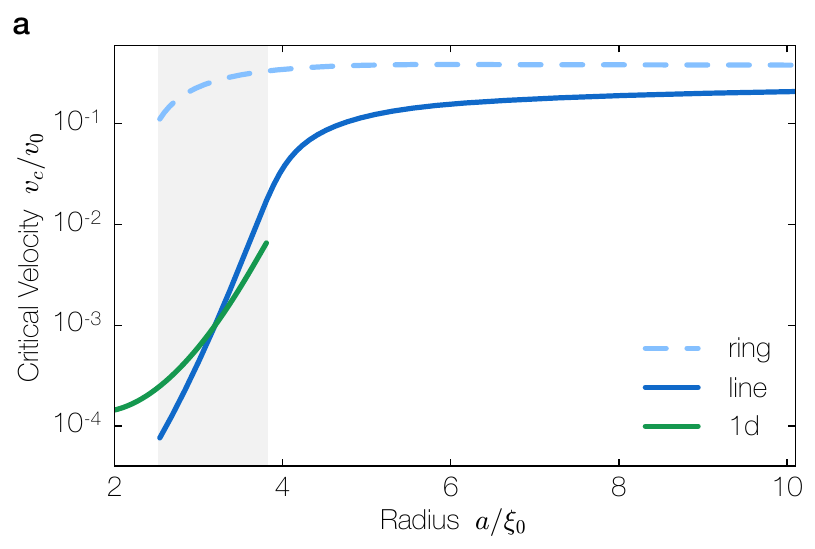}
\includegraphics[width=\columnwidth]{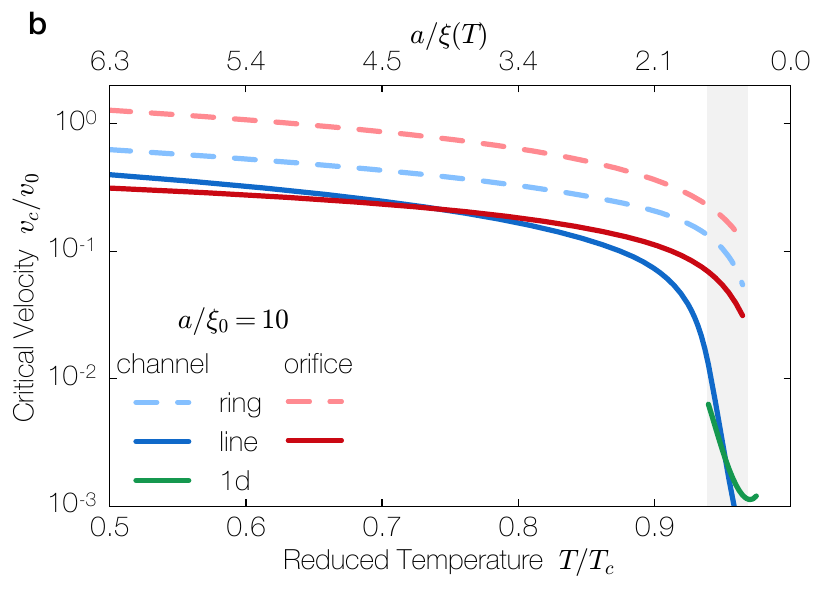}
\end{center}
\caption{
    \textbf{a} The critical velocity in a quasi-1d superfluid channel of length $L =
10^3\xi_0$ as a function of radius for $T/T_c = 3/4$,
$h \Gamma = k_{\rm B}T_c$ and $\xi(3T_c/4) \simeq 2.5 \xi_0$. The solid and dashed
blue lines extending over the full domain of the plot were computed via the
vortex nucleation theory, while the green line is for 1d,  
Eq.~\eqref{eq:Ex01d}.  \textbf{b} The temperature
dependence of the critical velocity for a channel ($L = 10^3 \xi_0$) and
orifice ($L = 10 \xi_0$) with $a = 10\xi_0$.  As $T\to T_c$, the
correlation length $\xi$ diverges, thus reducing the effective
channel width $a/\xi(T)$ and lowering $v_c$.
The shaded gray bars demarcates the radii where $1 \le
a/\xi(T) \le 3/2$ and in this region the upper bound due to ring vortices is no
longer expected to be relevant.}
\label{fig:velocityTemperature} 
\end{figure}

In the quasi-1d $(a \lesssim \xi)$ limit there are no transverse degrees of freedom, and the
system can be described in analogy to fluctuating superconducting wires by
computing the resistance due to thermally activated phase slips \cite{LA,MH}.
Translated into the language of 1d superfluidity, the phase slip
energy is (see supplemental material \cite{supplemental}):  
\begin{align}
\beta_c E_{1d} = \pi  \left({a \over \xi}\right)^2\left[{4 \over 3\sqrt{2}} 
\ - \  \pi {\xi \over \xi_0} { v_s  \over v_0}\right] .
 \label{eq:Ex01d}
\end{align}
For line vortices in the channel flow profile, the
tension scales with $(a/\xi)^2$ in contrast to the linear dependence in
Eq.~\eqref{eq:Ex0Line}. This difference is unimportant at the crossover 
$a/\xi \approx 1$ and the critical velocity from the 
nucleation of line vortices and the 1d theory should be in agreement as
demonstrated in Fig.~\ref{fig:velocityTemperature}a.  
%
% ------------------------------------------------------------------------------- 
% ------------------------------------------------------------------------------- 
%
The lower bound on the critical velocity due to line vortices drops three
orders of magnitude, and crosses over to the 1d result for single mode
channels.  An analogous crossover from the linear to non-linear Josephson
junction regime has been observed in superflow through an array of orifices
near $T_c$ \cite{hoskinson}.

Fig.~\ref{fig:velocityTemperature}b shows the $T$-dependence of $v_c$ 
for constrictions with $a/\xi_0 = 10$ 
and we observe a reduction as $T\to T_c$.  Qualitatively, this is due to the fact that 
$\xi(T)$ increases as $T\to T_c$, thus reducing the ratio
$a/\xi(T)$ which determines the effective constriction radius.  Experiments on
$^4$He have reported an apparent temperature power-law scaling of
$v_c$ \cite{duc} which we have confirmed is the spurious result of an interplay
between the thermal activation energy and vortex attempt rate. As $T\to 0$,
vortices will no longer be thermally activated and dissipation will be
dominated by the nucleation of quantum phase slips \cite{Khlebnikov:2004ia,
Danshita:2013di}.

% =============================================================================== 
% Discussion
% =============================================================================== 

In summary, we considered two characteristic confined flow geometries and
the thermal activation of representative low energy excitations, ring and
line vortices, inside them.  The resulting bounds they place on the critical velocity of
neutral superflow through narrow constrictions agree 
with a large body of measurements on confined superfluid $^4$He and
low-dimensional ultracold gases.  As the confinement radius approaches the
healing length, we find an exponential suppression of the critical velocity
of three orders of magnitude.  The experimental observation of this dramatic
reduction would be a clear signal of entering the strongly fluctuating
mesoscale regime.

% ----------------------------------------------------------------------------------
\noindent
\emph{Acknowledgements} --
We thank B.I. Halperin, G. Gervais, P.F. Duc, P. Taborek and K. Wright for helpful discussions.  A.D.
appreciates the University of Leipzig's hospitality and his participation
in a grand challenges workshop on quantum fluids and solids supported by NSF
DMR-1523582.  B.R. acknowledges support by DFG grant No. RO 2247/8-1.

%-----------------------------------
\input{main_refs}
%-----------------------------------

%%%%%%%%%% Merge with supplemental materials %%%%%%%%%%
\clearpage
\pagebreak
\onecolumngrid
\widetext
\begin{center}
\textbf{\large Supplementary material for ``Dissipation in mesoscale
superfluids''}
\end{center}
%%%%%%%%%% Merge with supplemental materials %%%%%%%%%%

%%%%%%%%%% Prefix a "S" to all counters and reset the counter %%%%%%%%%%
\setcounter{equation}{0}
\setcounter{figure}{0}
\setcounter{table}{0}
\setcounter{page}{1}
\makeatletter
\renewcommand{\theequation}{S\arabic{equation}}
\renewcommand{\thefigure}{S\arabic{figure}}
\renewcommand{\bibnumfmt}[1]{[S#1]}
\renewcommand{\citenumfont}[1]{S#1}
%%%%%%%%%% Prefix a "S" to all counters and reset the counter %%%%%%%%%%

\section{Velocity flow profiles}  
\label{sec:flowProfiles}

As illustrated in Fig.~1a-b of the main text, we
consider two different superfluid flow profiles through long narrow
\emph{channels}, and hole-like \emph{orifices}.

\subsection{Channel Flow}
\label{sub:Channel}
For a long cylindrical channel of radius $a$ and length $L \gg a$ oriented with
its axis along the $z$-direction we neglect any acceleration of the fluid at
the entry and exit to obtain a spatially independent velocity field
\begin{equation}
    \vec{v}_s = v_J \hat{z} = \frac{J}{\pi a^2 \rho_s} \hat{z} 
\label{eq:vConstant}
\end{equation}
where $J$ is the total mass flow rate and $\rho_s$ is the superfluid mass
density.

\subsection{Orifice Flow}
\label{sub:orifice}

The geometry of an orifice of radius $a$ oriented in the $x-y$ plane centered at the
origin can be conveniently described in oblate spheroidal coordinates
$(\zeta,\eta,\phi)$ by the surface $\eta = 0$ where
\begin{align}
    x &= a\cos\phi\sqrt{(1+\zeta^2)(1-\eta^2)} \notag \\
    y &= a\sin\phi\sqrt{(1+\zeta^2)(1-\eta^2)} \\
    z &= a\zeta \eta \notag \ .
\label{eq:oblateSphereCoords}
\end{align}
In general, $\phi \in [0,2\pi)$, $\zeta \in [0,\infty)$ and $\eta \in [-1,1)$ 
with the surface $|\eta| = \eta_0$ corresponding to a hyperboloid of revolution
about the $y$-axis (see Fig.~1b).  Steady incompressible
flow through the orifice with total rate $J$ can be studied by defining a potential field
$\Phi(\zeta,\eta,\phi)$ such
that $\vec{v}_s = (\hbar/m) \vec{\nabla} \Phi$ and solving $\nabla^2 \Phi = 0$
by requiring that $\Phi$ is continuous at $\zeta = 0$, and subject to the
boundary condition that the velocity component perpendicular to the surface
defined by $\eta = 0$ vanishes outside the orifice.  This yields the velocity
field: \cite{SImorsefeshbach,SIblount,SIschwarz}
\begin{equation}
\vec{v}_s = \frac{v_J}{2} \frac{1}{\sqrt{(\zeta^2+\eta^2)(\zeta^2 +
1)}}
    \hat{\zeta} 
\label{eq:vOrifice}
\end{equation}
which is plotted in the $y-z$ plane in Fig.~1c.
Inside the orifice at $\zeta = 0$ this expression simplifies to: 
\begin{equation}
    \vec{v}_s(r,z=0) = \frac{v_{J}}{2\sqrt{1-r^2/a^2}} \hat{z}
\label{eq:vOrificez0}
\end{equation}
which diverges near the boundary as $r\to a$.

% =============================================================================== 
\section{Vortex Energies}
\label{sec:vortex_energies}
% =============================================================================== 

Employing Eqs.~(1)--(2) in the main text in
combination with the spatial dependence of $v_s$ we may now obtain 
the energy cost of nucleating line and ring vortices.

% ------------------------------------------------------------------------------- 
\subsection{Channel Flow}
\label{sub:channel_flow}
% ------------------------------------------------------------------------------- 

For channels, we use the velocity profile in Eq.~\eqref{eq:vConstant}.

\subsubsection{Line Vortices}
\label{ssub:channel_flow_line_vortices}
\begin{figure}[h]
\begin{center}
\includegraphics[width=0.5\textwidth]{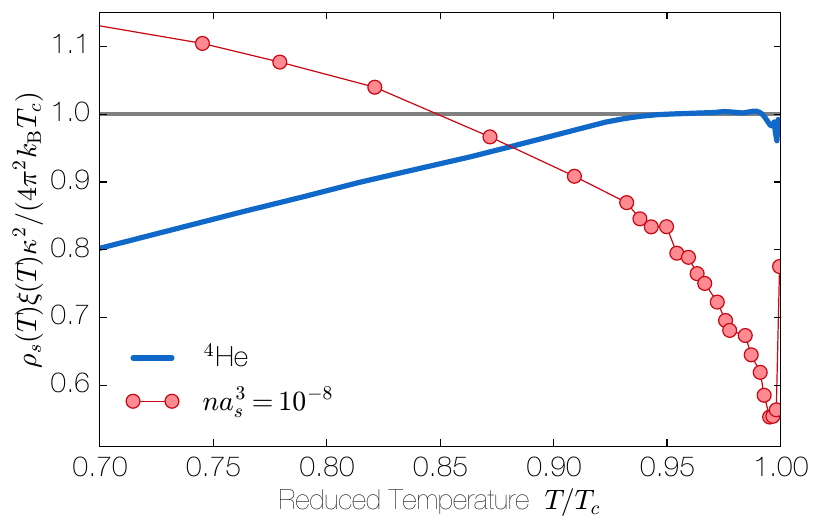}
\end{center}
\caption{The accuracy of the Josephson scaling relation defined in
Eq.~(\ref{josephson.eq}) compared with experimental results for the superfluid
density and correlation length at saturated vapor pressure from
Ref.~[\onlinecite{SIDonnelly:1998ge}] for $^4$He and theoretical
    results for a weakly interacting
Bose gas with $na^3 = 10^{-8}$  (data adapted from
Ref.~[\onlinecite{SIProkofev:2004js}]) where we have replaced the
scaling prefactor $\xi_0$ with the healing length $\xi_h$. See section III.B for
details. Here $n$ is the number density and $a_s$ the scattering length.}
\label{fig:js}
\end{figure}

% ............................................................................... 
For a line vortex attached to the walls of a channel of radius $a$ offset a
distance  $x$ from the center with length $\mathcal{L}(x) = 2\sqrt{a^2-x^2}$ as
seen in Fig.~\ref{vortextypes.fig} the 
total energy is the sum of $E_{\rm tension} + E_{\rm flow}$ and given  by:
\begin{align}
    E_{\rm line}(x) 
    &= \frac{\kappa^2\rho_s}{2\pi} \sqrt{a^2 - x^2}\left(\ln \frac{a-|x|}{\xi}
    + \alpha \right) \nonumber \\
    & \qquad - \kappa \rho_s v_J \left[a^2\cos^{-1}\left(\frac{x}{a}\right) -
x\sqrt{a^2 - x^2}\right]  .
\label{eq:ELineChannel}
\end{align}
Now, the superfluid density and correlation length $\xi$ can be related via 
the Josephson scaling relation in three dimensions \cite{SIjosephson}
\begin{equation}
    \xi(T) = \frac{4 \pi^2 k_{\rm B} T_c}{\kappa^2 \rho_s(T)}
    \label{josephson.eq}
\end{equation}
where $\xi(T) = \xi_0 (1-T/T_c)^{-\nu}$.  Fig.~\ref{fig:js} shows the accuracy
of this relation for superfluid $^4$He \cite{SIDonnelly:1998ge} and a weakly
interacting Bose gas \cite{SIProkofev:2004js} down to 
$T/T_c \approx 0.7$.
%
% ------------------------------------------------------------------------------- 
% ------------------------------------------------------------------------------- 
%
All temperature dependence now enters expressions through the correlation
length and the resulting dimensionless vortex energy is given by:
\begin{align}
    \beta_c E_{\rm line}(x) &= 2\pi\frac{a}{\xi}\sqrt{1 -
    \left(\frac{x}{a}\right)^2} \left[\ln  \left(1-\frac{|x|}{a}\right) + \ln
    \frac{a}{\xi} + \alpha \right] \nonumber \\
    &\!\! - \frac{v_J}{v_0}\left(\frac{a}{\xi}\right)^2\frac{\pi \xi}{\xi_0} \left[\cos^{-1}\left(\frac{x}{a}\right) - \frac{x}{a}\sqrt{1-\left(\frac{x}{a}\right)^2}\,\right] 
\label{eq:betaELineChannelSI}
\end{align}
with $\beta_c \equiv 1/(k_{\rm B}T_c)$ and we have identified the fundamental
velocity scale $v_0 = \kappa /(4\pi \xi_0)$.  This expression simplifies to
Eq.~(9)
%\eqref{main-eq:Ex0Line} 
in the main text when the vortex line is located at
the center of the channel $x=0$.

\subsubsection{Ring Vortices}
\label{ssub:channel_flow_ring_vortices}
% ............................................................................... 
For a ring vortex of radius $R$, with length $\mathcal{L} = 2\pi R$ the energy
can be written as:
\begin{equation}
    E_{\rm ring}(R) = \frac{1}{2} \kappa^2 \rho_s R \left(\ln \frac{R}{\xi} + \alpha \right)  - \kappa \rho_s \pi R^2 v_J \\
\label{eq:ERingChannel}
\end{equation}
yielding
\begin{equation}
    \beta_c E_{\rm ring}(R) = 2 \pi^2 \frac{R}{\xi} \left( \ln \frac{R}{\xi} + \alpha\right)
    - \frac{v_J}{v_0} \left(\frac{R}{\xi}\right)^2 \frac{\pi^2 \xi}{\xi_0}.
\label{eq:betaERingChannel}
\end{equation}
%

% ------------------------------------------------------------------------------- 
\subsection{Orifice Flow}
\label{sub:orifice_flow}
% ------------------------------------------------------------------------------- 
For the orifice flow profile, the velocity field now has the spatial dependence
seen in Eq.~\eqref{eq:vOrificez0} and Fig.~1c with a
divergence at the boundary.  While the effective line tension $E_{\rm tension}$
(first term) in Eqs.~\eqref{eq:ELineChannel} and \eqref{eq:ERingChannel} are
unchanged, a modified spatial integral in $E_{\rm flow}$ needs to be computed.

\subsubsection{Line Vortices}
\label{ssub:orifice_flow_line_vortices}
% ............................................................................... 

The energy cost for flow captured by a line vortex at position $x$ is given by
\begin{align}
    E_{\rm flow,line}(x) 
    &= \frac{\pi}{2}\kappa \rho_s v_J a^2 \left(1-\frac{x}{a}\right)
\label{eq:ELineOrifice}
\end{align}
leading to the total dimensionless vortex energy:
\begin{align}
    \beta_c E_{\rm line}(x)  &=  2  \pi  \frac{a}{\xi} \sqrt{1 - \left(\frac{x}{a}\right)^2}\left[\ln\left(1 - \frac{|x|}{a}\right) 
    + \ln \frac{a}{\xi} + \alpha \right] \nonumber \\
    & \qquad
    - \frac{v_J}{v_0} \left( \frac{a}{\xi} \right)^2 \frac{\pi^2 \xi}{2 \xi_0}
    \left(1-\frac{x}{a}\right)  .
\label{eq:betaELineOrificeSI}
\end{align}

\subsubsection{Ring Vortices}
\label{ssub:orifice_flow_ring_vortices}
% ............................................................................... 

For ring vortices, the flow integral is given by
\begin{equation}
    E_{\rm flow,ring}(R) = \kappa \rho_s v_J \pi a^2 \left[1-\sqrt{1-\left(\frac{R}{a}\right)^2}\, \right]
\label{eq:ERingOrifice}
\end{equation}
which leads to 
\begin{align}
    \beta_c E_{\rm ring}(R) &= 2 \pi^2 \frac{R}{\xi} \left( \ln \frac{R}{\xi} +
\alpha\right) \label{eq:betaERingOrificeSI} \\
& \quad    - \frac{v_J}{v_0} \left(\frac{a}{\xi}\right)^2 \frac{\pi^2 \xi}{\xi_0} 
    \left[1-\sqrt{1-\left(\frac{R}{a}\right)^2}\, \right] . \nonumber
\end{align}
%

% =============================================================================== 
\section{Application to Mass Flow Experiments}
\label{sec:mass_flow_experiments}
% =============================================================================== 
In this section we provide a more complete analysis of neutral bosonic mass
flow experiments that were discussed in Fig.~\ref{fig:exp_comparison_bec}
of the main text.  

% ------------------------------------------------------------------------------- 
\subsection{Superfluid Helium-4}
\label{sub:superfluid_helium_4}
% ------------------------------------------------------------------------------- 

\begin{figure}[t]
\begin{center}
    \includegraphics[width=0.5\textwidth]{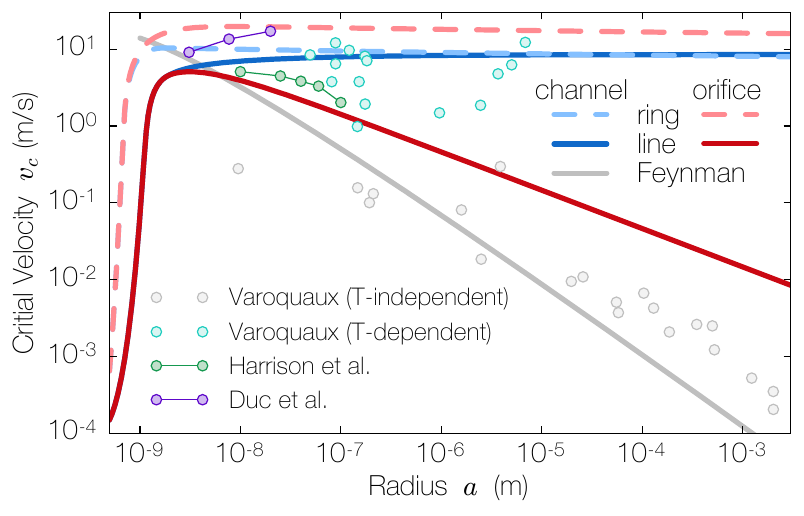}
\end{center}
\caption{
A compilation of experimental results on the critical superfluid velocity $v_c$
of $^4$He under pressure driven flow through constrictions with radius $a$ from
Varoquaux (temperature independent and dependent) \cite{SIvaroquauxrmp}, Harrison
\emph{et al.} \cite{SIharrison} and Duc \emph{et al.} \cite{SIduc}.  The
temperature independent results for larger channels are qualitatively described
by the Feynman critical velocity  $v_{c,F} = (\kappa/4\pi a)\ln(2a/\xi_0)$ where
$\xi_0 \simeq \SI{3.45}{\angstrom}$ is the zero temperature coherence length.
For smaller pores with larger temperature dependent critical velocities, the
data is well-bounded by the predictions of the vortex nucleation theory for
ring and line vortices in the orifice and channel flow profiles with $L =
\SI{30}{\nano\meter}$ and $\Gamma = \SI{4}{\giga\hertz}$ at
$T=\SI{1.5}{\kelvin}$, values consistent with those in the experiment \cite{SIduc}, and representative for the other experiments.} 
\label{fig:exp_comparison}
\end{figure}

A recent review by Varoquaux \cite{SIvaroquauxrmp} includes a compilation of
superfluid critical velocity results from a diverse set of experiments (both
dependent and independent of temperature) that are reproduced in
Fig.~\ref{fig:exp_comparison} along with theoretical upper and lower bounds
computed within the vortex nucleation theory using parameters relevant for
superfluid helium.  The experimental data from Ref.~[\onlinecite{SIvaroquauxrmp}]
can be grouped into two distinct regions. For larger channels, a temperature
independent critical velocity increases as the channel size is reduced (grey
circles).  
%
% ------------------------------------------------------------------------------- 
% ------------------------------------------------------------------------------- 
%
This is consistent with the Feynman prediction $v_F \sim \kappa/a$, where
vortex rings are dynamically ejected from the end of the channel.

%
% ------------------------------------------------------------------------------- 
\begin{table*}
\begin{center}
    \renewcommand{\arraystretch}{1.6}
    \setlength\tabcolsep{2ex}
  \begin{tabular}{r c c c c c c}
   \hline\hline 
   \textbf{Reference} & \textbf{System} & $v_c$
   (\si{\milli\meter\per\second}) & $v_0$ (\si{\milli\meter\per\second}) & 
   $a$ (\si{\micro\meter}) & $\xi_0$ (\si{\micro\meter}) & $T/T_c$ \\
   \hline
   Neely \emph{et al.} [\onlinecite{SIneely}] & $^{87}$Rb & 0.2 & 1.2 & 47 &  0.3
   & 0.6\\
   Ramanathan \emph{et al.} [\onlinecite{SIramanathan}] & $^{23}$Na & 0.9 & 0.7 &
   7 & 0.51 & 0.2\\
   Raman \emph{et al.} [\onlinecite{SIraman}] & $^{23}$Na & 1.6 & 4.6 & 15 & 0.3
   & 0.8\\
   Weimer \emph{et al.} [\onlinecite{SIweimer}] & $^{6}$Li (molecule) & 1.7 & 3.1
   & 10 & 0.85 & 0.5\\
    \hline \hline
  \end{tabular}
\end{center}
\caption{\label{tab:BECData} Critical velocity and related velocity, length
and temperature scales extracted from four neutral bosonic mass flow experiments in ultracold
atomic and molecular condensates.}
\end{table*}
% ------------------------------------------------------------------------------- 
%

A second group of temperature dependent critical velocities in pores with radii
$a\sim \SI{100}{\nano\meter} - \SI{10}{\micro\meter}$ have considerably larger
velocities and are in a thermally activated dissipation regime that is well
bounded by the vortex nucleation theory.  Additionally,
Fig.~\ref{fig:exp_comparison} includes a systematic set of experiments performed by
Harrison and Mendelssohn \cite{SIharrison} (green circles) employing the fountain
effect to drive superfluid mass flow through arrays of $10^4-10^7$
$L=\SI{5}{\micro\meter}$ pores etched in irradiated mica using HF acid with
radii $a = 20,50,80,120$ and $\SI{200}{\nano\meter}$.  These results, taken at
$T = \SI{1.5}{\kelvin}$, should be considered as representing a lower bound
on the critical velocity as all pores were assumed to be open in the analysis
and both a variation in radii and a taper along the channel were observed.

Experiments by Duc \emph{et al.} \cite{SIduc} shown in
Fig.~\ref{fig:exp_comparison} are in the interesting mesoscopic regime where
$a/[\xi_0 (1-T/T_c)^{-\nu}] \sim \mathrm{O}(1)$. For superfluid helium, $\xi_0
\simeq \SI{3.45}{\angstrom}$, $T_c = T_{\lambda} \simeq \SI{2.1768}{\kelvin}$
and $\nu \simeq 0.6717$.  They observed a decrease in the critical velocity as
the pore radius was reduced.  In these experiments, $^4$He mass flow is studied
through single pores nanofabricated using a transmission electron beam incident
on a $L \simeq \SI{30}{\nano\meter}$ thick silicon nitride wafer resulting in
smooth constrictions with radii $a \simeq
\text{\SIlist{3.14;7.81;20}{\nano\meter}}$.  Critical velocities were reported
for $T = \SI{1.5}{\kelvin}$, and flow was driven via pressure differences
between $\Delta P= \SIrange{250}{830}{\milli\bar}$ which corresponds to an
external drive frequency $\Gamma \simeq \SIrange{2}{7}{\giga\hertz} \sim k_{\rm
B} T_c / h$.  

% ------------------------------------------------------------------------------- 
\subsection{Ultracold Gases}
\label{sub:becs}
% ------------------------------------------------------------------------------- 

In systems of neutral ultracold atoms,  flow is driven by dynamically 
shaping an optical potential which causes a local imbalance in the
chemical potential (see Ref.~[\onlinecite{SIchien}] for current experimental
designs).  
For dilute gases, the prefactor $\xi_0$ which appears in the critical scaling
relation for the correlation length can be related to the zero temperature
healing length $\xi_h = 1/\sqrt{8\pi a_s n}$ via known universal results for
the weakly interacting Bose gas \cite{SIProkofev:2004js}.  Here $n$ is the
number density and $a_s$ is the
scattering length and provided that $na^3 \ll 10^{-5}$ the superfluid density
can be written in terms of a universal scaling function $f_s$:
\begin{equation}
    \rho_s(t) = \frac{16 \pi^3 m n}{\left[\zeta\left(\frac{3}{2}\right)\right]^{4/3}} 
    (na^3)^{1/3} \left(1-t\right)^2 f_s\left(t,na^3\right) 
\label{eq:rho_sWIBG}
\end{equation}
where $m$ is the mass, $t = 1-T/T_c$ is the reduced temperature
and the Riemann zeta function, $\zeta(3/2)$, appears through the use of
the critical temperature of the non-interacting Bose gas: $k_{\rm B}T_c =
({2\pi\hbar^2}/{m}) [{n}/{\zeta(3/2)}]^{2/3}$. The function $f_s$ is known from
high precision Monte Carlo calculations
\cite{SIProkofev:2004js,SICapogrossoSansone:2010kt} and when $t\ll1$, $f_s \propto
(t/\sqrt[3]{na^3})^\nu$. Writing the temperature dependent correlation length
in the critical region as $\xi(T) = (\xi_0/\xi_h) \xi_h t^{-\nu}$, we can use
the Josephson relation in Eq.~\eqref{josephson.eq} to determine
\begin{equation}
    % \frac{\xi_0}{\xi_{h}} = \mathcal{A} \left\{\frac{32 \pi^3 \left(na^3\right)^{1/3}}{\left[\zeta\left(\frac{3}{2}\right)\right]^{4/3}} \right\}^{\nu-1/2}
\frac{\xi_0}{\xi_{h}} = \mathcal{A}  \left(na^3\right)^{(2\nu-1)/6} 
%\approx \mathcal{A} \left(na^3\right)^{1/18} 
\label{eq:xi0oxih}
\end{equation}
where $\mathcal{A} \simeq 3.8$ is a universal number. Thus, for experimentally
accessible weakly interacting Bose gases $\xi_0 \simeq \mathcal{B} \xi_h$ with
$\mathcal{B} \sim 1-2$. Away from the critical region, the full scaling
function can be used to test the accuracy of the Josephson relation with the
result shown in Fig.~\ref{fig:js}.

Thus, to analyze non-equilbirium mass flow experiments employing ultracold
Bose-Einstein condensates, we approximate $\xi_0 \approx \xi_h \sim
\SI{1}{\micro\meter}$, which yields
the critical velocity scale $v_0 = h/(4\pi m \xi_0) \sim
\SI{1}{\milli\meter\per\second}$. 
In Fig.~\ref{fig:exp_comparison_bec} of the main text, we have shown
results for the critical velocity from three ultracold gas experiments using the
parametrization shown in Table~\ref{tab:BECData}. We have not included the low
temperature data point from Ref.~[\onlinecite{SIramanathan}] where the
errors in our critical theory are difficult to estimate.
We find that the other experiments are in the
same flow regime as tightly confined superfluid helium. 

% =============================================================================== 
\section{Intrinsic critical velocity of bulk superfluids}
\label{app:intrinsic_vc}
% =============================================================================== 

In the $a \gg \xi_0$ limit we can directly find the critical ring
vortex radius $R^\ast$ that maximizes the energy barrier in the constant
channel flow profile in Eq.~\eqref{eq:ERingChannel} as
\begin{equation}
\frac{R^\ast}{\xi} = \frac{v_0}{v_J} \frac{\xi_0}{\xi} \left(
\ln \frac{R^\ast}{\xi}+\alpha + 1\right) .
\label{eq:Rcritical}
\end{equation}
In this bulk regime, one only needs to consider the effects of vortices which
reduce the total energy and Eq.~(5)
%\eqref{main-eq:decayRate} 
 gives
\begin{align}
    \ln \frac{\Gamma_0} {\Gamma} &= \beta E_{\rm ring}(R^\ast) \label{eq:lnGamma} \\
    &= \frac{T_c}{T} \left[2\pi^2 \frac{R^\ast}{\xi}
\left( \ln \frac{R^\ast}{\xi}+\alpha \right)  -
\frac{v_c}{v_0}\left(\frac{R^\ast}{\xi}\right)^2 \frac{\pi^2 \xi}{\xi_0}\right]
  \nonumber  
\end{align}
which can be solved for the critical velocity $v_c$.  Combining
Eq.~\eqref{eq:lnGamma} with Eq.~\eqref{eq:Rcritical} with $v_J=v_c$ gives the
transcendental equation
\begin{equation}
    \frac{R^\ast}{\xi} = 2\frac{\ln \frac{R^\ast}{\xi}+\alpha  -
    \frac{1}{2\pi^2}\frac{T}{T_c} \ln \frac{\Gamma_0}{\Gamma} } 
    {  \ln \frac{R^\ast}{\xi}+\alpha +1 }
\label{eq:RcritTrans}
\end{equation}
which can be simplified to yield Eq.~(7).
%\eqref{main-eq:RcritEqn}.  
Replacing $\ln
\Gamma_0/\Gamma$ in Eq.~\eqref{eq:lnGamma} via Eq.~\eqref{eq:RcritTrans} we
obtain the critical velocity in the large radius limit in
Eq.~(8).
%\eqref{main-eq:vcEqn}.

\begin{figure}[t]
\begin{center}
    \includegraphics[width=0.5\textwidth]{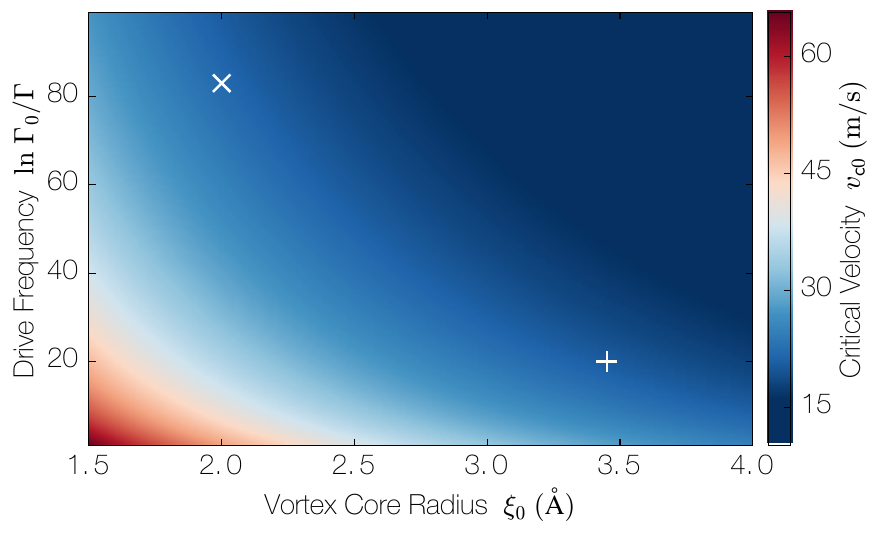}
\end{center}
\caption{The intrinsic critical velocity scale $v_{c0}$ of superfluid $^4$He as
    a function of the effective zero temperature core size $\xi_0$ and external drive
    frequency $\ln \Gamma_0/\Gamma$. In order to obtain a physically meaningful
    value of the critical velocity $v_c = v_{c0}(1-T/T_c)^\nu$ when using $\xi_0 =
    \SI{2}{\angstrom}$, Langer and Fisher \cite{SIlanger} had to consider
an extremal value for the external drive as indicated by the cross
($\times$).  When utilizing a larger core size of $\xi_0=\SI{3.45}{\angstrom}$
as experimentally determined in Ref.~[\onlinecite{SISiAh84}], $\ln
\Gamma_0/\Gamma$ can be reduced to more physically meaningful values as
indicated by the plus $(+)$.} 
\label{fig:LFVelocity}
\end{figure}

Eq.~(7)
%\eqref{main-eq:RcritEqn} 
in the main text is equivalent to Eq.~(13) in
Langer and Fisher \cite{SIlanger} (LF), noting the change of variables $\eta_c
\equiv \ln 8R^\ast/\xi$, their use of a rigid vortex core with $\alpha = \ln 8
- 7/4$ and setting $\ln \Gamma_0/\Gamma \simeq 83$.  The un-physically large
value of $\ln \Gamma/\Gamma_0$ employed by LF was required to ensure that their
computed value of $v_c = v_{c0}(1-T/T_c)^\nu$ in the scaling regime was below
the absolute upper bound given by the Landau criterion.  In order to understand
the origin of this long-standing discrepancy, we have investigated $v_{c0}$ as
a function of both the external drive $\Gamma_0/\Gamma$ and vortex core details
$\xi_0$ with the results shown in Fig.~\ref{fig:LFVelocity}.

%
% ------------------------------------------------------------------------------- 
% ------------------------------------------------------------------------------- 
%

% =================================================================================
\section{The one-dimensional limit}
\label{app:LAMH}
% =================================================================================

To derive Eq.~(11) in the main text, we begin with a free energy functional
near $T_c$:
\begin{equation}
\label{eq:freeEnergyDensity}
f = f_0 +\frac{\hbar^2}{2 m} \lvert \nabla \Psi \rvert^2 + A \lvert \Psi
\rvert^2 + \frac{B}{2} \lvert \Psi \rvert^4 
\end{equation}
where $f_0$ is a condensate energy density and the wave function is
given by: 
$
\Psi(\vec{r}) = \sqrt{n(\vec{r})} \mathrm{e}^{i \Phi(\vec{r})}
$
with $n(\vec{r})$ the number density.  In a spatially
homogeneous superfluid, the free energy is minimized when $\lvert \Psi \rvert^2
\equiv n_0 = - A/B$.  Substituting this value for the field into  Eq.
\eqref{eq:freeEnergyDensity}:
\begin{equation}
f - f_0 = - \frac{A^2}{2 B} = -\frac{H_c^2}{8\pi} \ , 
\end{equation}
where the last equality is schematic and in analogy to superconductors 
where $H_c$ is the critical field.  Identifying
$-A/B$ with the density and performing the usual rescaling:
$\tilde{\Psi} = \Psi/\sqrt{n_0}$ one introduces the correlation 
length $\xi(T)$ such that
$\xi^2(T) = \hbar^2/2 m A(T)$, 
and we can now write
$ f -  f_0 = \rho_s \kappa^2/16\pi^2 \xi^2(T) $.
According to Refs.~[\onlinecite{SILA}-\onlinecite{SIMH}], the free energy barrier of a phase slip excitation 
inside a quasi-one-dimensional channel of cross-sectional area $\pi a^2$ is
given by 
%
%***************** LAMH energy barrier  ****************
\begin{equation}
\Delta F_0 = \frac{8 \sqrt{2}}{3} \frac{H_c^2}{8\pi} A \xi = \frac{\sqrt{2}}{6\pi} \rho_s \kappa^2 \xi \left(\frac{a}{\xi}\right)^2 \ \ .
\label{eq:DF0}
\end{equation}
%*******************************************************
%
The energy reduction due to a superflow $J$ is $ \delta F = \kappa \rho_s v_J
\pi a^2$, and adding the free energy barrier  Eqs.~\eqref{eq:DF0} one obtains
the total phase slip energy in Eq.~(11) of the main text. The total phase slip
rate is
$\Gamma = 2 \Gamma_0 \mathrm{e}^{-\Delta F_0/k_{\rm B} T} \sinh (\delta F/k_{\rm B} T)$,
and solving for $v_c$ we find
\begin{align}
\label{eq:vs1d}
\frac{v_c}{v_0} &= \frac{1}{\pi} \left(\frac{\xi_0}{a}\right) {T \over T_c} \left(1 - {T \over T_c}\right)^{-\nu} 
\\
& \quad \times \sinh^{-1} \left[\frac{\Gamma}{2\Gamma_0} \exp\left(
{4 \pi \over 3 \sqrt{2}} {a^2 \over \xi_0^2} {T_c \over T} \left(1 - {T \over T_c}\right)^{2 \nu}
\right)\right] , \notag
\end{align}
where \cite{SIMH}
\begin{align}
\label{eq:Gamma0}
\Gamma_0 
&=  {a L \over \xi_0^2} \left(1 - {T \over T_c}\right)^{2 \nu +1} \left(\frac{2\sqrt{2}\pi }{3} \frac{T_c}{T}\right)^{1/2}
{8 k_B T_c   \over \pi \hbar } \ \ .
\end{align}
Eq.~\eqref{eq:vs1d} 
is shown in Fig.~\ref{fig:velocityTemperature} 
of the main text.
% ---------------------------------------------------------------------------------

%-----------------------------------
\input{supp_refs}
%-----------------------------------

\end{document}